\pdfoutput=1
\documentclass{article}
\usepackage{spconf,amsmath,amsfonts,mathrsfs}
\usepackage[retainorgcmds]{IEEEtrantools}
\usepackage{cite}
\usepackage[ruled,longend]{algorithm2e}
\usepackage{graphicx}
\usepackage{caption}
\usepackage{subcaption}

\title{Spectral Compressive Sensing with Model Selection}
\name{Zhenqi~Lu, Rendong~Ying, Sumxin~Jiang, Zenghui~Zhang, Peilin~Liu, Wenxian~Yu
\thanks{This work was partially supported by the National Natural Science Foundation of China under grant number 61171171 and 61102169. }}
\address{Dept.~of~Electrical~Engineering,~Shanghai~Jiao~Tong~University,~Shanghai,~P.~R.~China}
\begin{document}
%\ninept
%
\maketitle
\begin{abstract}
 The performance of existing approaches to the recovery of frequency-sparse signals from compressed measurements is limited by the coherence of required sparsity dictionaries and the discretization of frequency parameter space. In this paper, we adopt a parametric joint recovery-estimation method based on model selection in spectral compressive sensing. Numerical experiments show that our approach outperforms most state-of-the-art spectral CS recovery approaches in fidelity, tolerance to noise and computation efficiency.
\end{abstract}
\section{Introduction}
\label{sec:introduction}
One of the recent research interests of compressive sensing (CS) has focused on the recovery of signals that are spectrally sparse from a reduced number of measurements \cite{candes2007sparsity,tropp2010beyond,mishali2010theory,wakin2012nonuniform,duarte2013spectral,duarte2011structured}. A great many applications, including spectrum sensing\cite{meng2011collaborative} and wideband communication\cite{mishali2011xampling,mishali2010theory}, feature smooth or modulated signals that can be modelled as a superposition of a small number of sinusoids. Recovery of such \emph{frequency-sparse signals} brings about a novel issue in the formulation of CS recovery problem: signal representations in frequency domain have a continuous parameter space, while recent CS researches\cite{donoho2006compressed,candes2006near,baraniuk2007compressive} are rooted on signal decomposition in a discretized dictionary.
\newline\indent An intuitive solution to this problem is a denser sampling of the parameter space, which improves the compressibility of signal representations. But increasing the resolution of parameter sampling worsens the coherence between dictionary elements, which results in loss of sparsity and uniqueness of signal representations. Such ambiguity prevents certain algorithms \cite{donoho2006compressed,candes2006near} from achieving the sparse representation successfully. Initial contributions to spectral CS recovery are concentrated on the recovery algorithm, the optimization problem formulation and the sparsity prior to combat the intricacy in signal representations \cite{candes2011compressed,duarte2013spectral,tang2012compressed,fyhn2013spectral,fannjiang2012coherence,ekanadham2011recovery}.
\newline\indent Estimation of the frequencies, amplitudes and phase shifts of sinusoids embedded in noise is a fundamental problem in time series analysis and other more general statistical signal processing problems. Classical treatments of this problem are restricted to the Fourier frequencies. This implicit discretization barrier is overcome due to the introduction of the \emph{minimum description length} (MDL) principle for \emph{model selection} \cite{nadler2011model,hannan1993determining,kavalieris1994determining}. In this paper, we improve over existing approaches by applying model selection to spectral CS. We take into consideration \emph{parametric} joint recovery-estimation methods, which determine parameters by minimizing a log-likelihood function of compressed measurements.
\newline\indent The novelty of our approach is that it performs estimation from compressed measurements rather than from signal samples. Since the log-likelihood minimization is in practice equivalent to an \smash{$\ell_2$}-norm minimization, the estimation performance is guaranteed as long as the sensing matrix satisfies \emph{restricted isometry property} (RIP), and thus distance-preserving \cite{candes2006near}. It can be shown that random marices from Gaussian, Rademacher, or more generally a sub-Gaussian distribution have the RIP with high probability under certain conditions \cite{baraniuk2008simple}. We solve the optimization problem through an iterative greedy approach to outperform state-of-the-art spectral CS approaches. Experimental results show improved reconstruction fidelity against existing approaches, from both noiseless and noisy measurements. In addition, our approach is essentially greedy, and is thus more computationally efficient than optimization based approaches. Furthermore, compared to traditional model selection estimators\cite{nadler2011model,hannan1993determining,kavalieris1994determining}, our approach reduces the number of samples needed and consequently the computational load of estimation.

\section{Problem Formulation and Related Prior Work}
\label{sec:problemform}
The problem of recovering frequency-sparse signals from compressed measurements is formulated as follows: Let \smash{$s\left(t\right)$} denote an unidimensional real-valued frequency-sparse signal composed of \smash{$K$} sinusoids with unknown frequencies \smash{$\omega_j$}, amplitudes \smash{$a_j$}, and phase shifts \smash{$\phi_j$}, and \smash{$x\left(t\right)$} be \smash{$s\left(t\right)$} corrupted by additive noise \smash{$\xi\left(t\right)$} with unknown noise level.
\begin{equation}\label{signalmodel}
    x\left(t\right) = s\left(t\right) + \xi\left(t\right) = \sum\limits_{j=1}^K a_j \sin\left(\omega_j t+\phi_j\right) + \xi\left(t\right).
\end{equation}
\noindent Let \smash{$\boldsymbol{s}=\left\lbrace s_t\right\rbrace_{t=1}^N$} and \smash{$\boldsymbol{x}=\left\lbrace x_t \right\rbrace_{t=1}^N$} be the observed samples of \smash{$s\left(t\right)$} and \smash{$x\left(t\right)$} at discrete times \smash{$\left\lbrace t_j=j \right\rbrace_{j=1}^N$}, and \smash{$\boldsymbol{m} = \boldsymbol{\Phi} \boldsymbol{x} \in \mathbb{R}^M$} be the compressed measurements of \smash{$\boldsymbol{x}$}, where $\boldsymbol{\Phi}\in\mathbb{R}^{M\times N}$ is the sensing matrix.
\newline\indent Given the compressed measurements \smash{$\boldsymbol{m}$}, the problem is to recover the \smash{$K$} sinusoids in \eqref{signalmodel}. Many different approaches have been suggested in the literature for spectral CS recovery. Under certain conditions for matrix \smash{$\boldsymbol{\Phi}$}, one can recover signal \smash{$\boldsymbol{s}$} through an \smash{$\ell_1$}-norm optimization problem \cite{donoho2006compressed,candes2006near}, denoted as \smash{$\ell_1$}-\emph{synthesis},
\begin{equation}\label{eqn:tradition}
    \boldsymbol{\hat{s}} =\boldsymbol{\mathrm{F}}\boldsymbol{\hat{\eta}},\boldsymbol{\hat{\eta}} = \arg\min_{\boldsymbol{\eta} \in \mathbb{R}^N} \left\lVert \boldsymbol{\eta} \right\rVert_1\ \text{s.t.}\ \left\lVert\boldsymbol{m}-\boldsymbol{\Phi} \boldsymbol{\mathrm{F}}\boldsymbol{\eta}\right\rVert_2 \leq \epsilon,
\end{equation}
\noindent where \smash{$\epsilon$} is an appropriately chosen bound on the noise level, \smash{$\boldsymbol{\mathrm{F}}$} is the orthonormal DFT basis, and \smash{$\boldsymbol{\eta}$} is the DFT coefficients of signal samples \smash{$\boldsymbol{s}$}. The optimal recovery of signal \smash{$\boldsymbol{s}$} by optimizing \eqref{eqn:tradition} is feasible provided that the decomposition of \smash{$\boldsymbol{s}$} in the DFT basis \smash{$\boldsymbol{\mathrm{F}}$} is \smash{$K$}-sparse, i.e. \smash{$\left\lVert\boldsymbol{\eta}\right\rVert_0 = K$} \cite{donoho2006compressed,candes2006near}. Unfortunately, not only the DFT coefficients of frequency-sparse signals are \emph{not sparse}, but even worse, they are \emph{just barely compressible}. One way to remedy this problem would be to employ a redundant DFT frame
\begin{IEEEeqnarray}{rCl}
    \boldsymbol{\Psi}\left(c\right) & := & \left\lbrack\boldsymbol{e}\left(0\right)\ \boldsymbol{e}\left(\Delta\right) \cdots \boldsymbol{e}\left(2\pi-\Delta\right)\right\rbrack,\Delta := 2\pi / cN,\IEEEnonumber\\
    \boldsymbol{e}\left(\omega\right) & := & \frac{1}{\sqrt{N}}\left\lbrack 1\ e^{j\omega} e^{j2\omega} \cdots e^{j\omega\left(N-1\right)}\right\rbrack^T,
\end{IEEEeqnarray}
\noindent as a substitute for \smash{$\boldsymbol{\mathrm{F}}$}. But the redundant DFT frame violates the incoherence requirement for the dictionary \cite{duarte2013spectral}.
\newline\indent It has recently been shown that the incoherence condition of dictionary \smash{$\boldsymbol{\mathrm{D}}$} is not necessary concerning the recovery of signal \smash{$\boldsymbol{x}$}, provided that the frame coefficients \smash{$\boldsymbol{\mathrm{D}}^H\boldsymbol{x}$} are sufficiently sparse \cite{candes2011compressed}, where \smash{$\left(\cdot\right)^H$} designates the Hermitian operation. Under this circumstance, \smash{$\ell_1$}-analysis yields good recovery result for signal \smash{$\boldsymbol{x}$}. However, the redundant DFT frame coefficients of frequency-sparse signals \smash{$\boldsymbol{\Psi}\left(c\right)^H\boldsymbol{s}$} do not have the sparsity property.
\newline\indent An alternative approach is to benefit from structured sparsity by using a coherence inhibition signal model \cite{duarte2013spectral}. The resulting Structured Iterative Hard Thresholding (SIHT) algorithm is able to recover frequency-sparse signals by selecting elements with low coherence in an redundant DFT frame. Other algorithms with similar flavor to SIHT include Band-excluded Orthogonal Matching Pursuit (BOMP), which takes advantage of band-exclusion \cite{fannjiang2012coherence}. However, the reconstruction fidelity of SIHT and BOMP is substantially limited due to the simplicity in the formulation of algorithms.
\newline\indent One way to remedy the discretization of parameter space is the \emph{polar interpolation} approach, and the corresponding algorithm is named Continuous Basis Pursuit (CBP) \cite{ekanadham2011recovery}. Like other optimization based algorithms, CBP suffers from its high computational complexity. A novel algorithm, Band-excluded Interpolating Subspace Pursuit (BISP), combining the merits of band-exclusion and polar interpolation, has been proposed more recently \cite{fyhn2013spectral,fyhn2013compressive}. By incorporating polar interpolation with greedy algorihtm, BISP improves the convergence rate of CBP while only inducing an amenable reduction in performance.
\newline\indent Recent advances in convex geometry has proved that frequency-sparse signals can be recovered from random subsamples via \emph{atomic norm minimization} \cite{tang2012compressed}, which can be implemented as a semidefinite program (SDP)\cite{chandrasekaran2012convex}. Though atomic norm has several appealing properties, SDP is in practice computationally expensive. Moreover, the formulation of SDP is limited to random subsampling matrix, and no discussion for arbitrary measurement settings is provided.

\section{Model Selection for Spectral Compressive Sensing}
\label{sec:modelselection}
\indent In this paper, we adopt a parametric joint recovery-estimation method, which estimates the unknown frequencies, amplitudes, and phase shifts. Under the assumption of white Gaussian noise, with the number of sinusoids \smash{$K$} \emph{a priori} known, a common method to estimate the \smash{$3K$} parameters is by maximizing the likelihood function \smash{$\mathcal{L}$} of observed data \smash{$\boldsymbol{x}$} \cite{nadler2011model}
\begin{equation}
    \mathcal{L}\left(\boldsymbol{\theta}_K,\boldsymbol{x}\right)=\prod_{t=1}^N e^{-\left\vert x_t - \sum_{j=1}^K a_j \sin\left(\omega_j t + \phi_j\right)\right\vert^2},
\end{equation}
\noindent where \smash{$\boldsymbol{\theta}_K = \left\lbrace a_j,\omega_j,\phi_j\right\rbrace_{j=1}^K$} contains the \smash{$3K$} parameters of the \smash{$K$} sinusoids. In spectral CS recovery problem, the signal samples \smash{$\boldsymbol{x}$} is not available, and as a substitute the compressed measurements \smash{$\boldsymbol{m}$} is observed. Thus the estimation method is reformulated as the minimization of the log-likelihood function of \smash{$\boldsymbol{m}$}
\begin{equation}
     \boldsymbol{\hat{\theta}}_K= \arg\min_{\boldsymbol{\theta}_K}-\ln\mathcal{L}\left(\boldsymbol{\theta}_K,\boldsymbol{m}\right),
\end{equation}
\noindent which is equivalent to an \smash{$\ell_2$}-norm minimization problem
\begin{equation}\label{eqn:ell2mini}
     \boldsymbol{\hat{\theta}}_K = \arg\min_{\boldsymbol{\theta}_K} \left\lVert \boldsymbol{m} - \sum_{j=1}^K\boldsymbol{\Phi}\boldsymbol{s}_j  \right\rVert_2^2,
\end{equation}
\noindent where \smash{$\boldsymbol{s}_j = \left\lbrace a_j \sin \left(\omega_j t + \phi_j\right)\right\rbrace_{t=1}^N$} is the sample vector of \smash{$K$} sinusoids with estimated \smash{$3K$} parameters. Obviously, \eqref{eqn:ell2mini} can be reformulated as the following \smash{$\ell_2$}-norm minimization problem
\begin{equation}\label{eqn:ell2mini:reform}
    \boldsymbol{\hat{\vartheta}}_K = \arg\min_{\boldsymbol{\vartheta}_K} \left\lVert\boldsymbol{m} - \sum_{j=1}^K\boldsymbol{\Phi}\left(a_{1,j}\boldsymbol{\sin}_{\omega_j}+a_{2,j}\boldsymbol{\cos}_{\omega_j}\right)\right\rVert_2^2,
\end{equation}
\noindent where \smash{$\boldsymbol{\vartheta}_K = \left\lbrace\omega_j,a_{1,j},a_{2,j}\right\rbrace_{j=1}^K$} contains the reformulated \smash{$3K$} parameters with \smash{$a_{1,j} = a_j\cos\phi_j$, $a_{2,j} = a_j\sin\phi_j$} designating the amplitudes of sine and cosine sinusoids, and \smash{$\boldsymbol{\sin}_{\omega_j}=\left\lbrace\sin n\omega_j\right\rbrace_{n=1}^N$}, \smash{$\boldsymbol{\cos}_{\omega_j}=\left\lbrace\cos n\omega_j\right\rbrace_{n=1}^N$} are the samples of sinusoids with frequency \smash{$\omega_j$}.
\newline\indent In recent parametric estimation works, the best \smash{$K$} matching sinusoids are iteratively recovered \cite{nadler2011model}. In this paper, we reformulate previous methods as a spectral CS recovery method as shown in Algorithm \ref{algorithm:model}, which deviates from previous approaches in that the input is compressed measurements rather than signal samples. In each iteration, the estimated compressed measurements of other \smash{$K-1$} sinusoids are trimmed from input compressed measurements \smash{$\boldsymbol{m}$}, and then the parameters of the best matching sinusoid to the residual measurements \smash{$\boldsymbol{r}$} are estimated through function \smash{$\mathcal{R}\left(\boldsymbol{\Phi},\boldsymbol{r}\right)$}.
\newline\indent The function \smash{$\lbrace\hat{\omega},\hat{a},\hat{\phi}\rbrace=\mathcal{R}\left(\boldsymbol{\Phi},\boldsymbol{r}\right)$}, as shown in Algorithm \ref{algorithm:parameter}, estimates the parameters \smash{$\boldsymbol{\hat{\theta}}=\lbrace\hat{\omega},\hat{a},\hat{\phi}\rbrace$} of the best matching sinusoid from residual CS measurement vector \smash{$\boldsymbol{r}$}. The algorithm solves the parametric estimation by minimizing the following log-likelihood function
\begin{equation}\label{eqn:likelihood:single}
    \boldsymbol{\hat{\theta}} = \arg\min_{\boldsymbol{\theta}}-\ln\mathcal{L}\left(\boldsymbol{\theta},\boldsymbol{r}\right),
\end{equation}
which is equivalent to an \smash{$\ell_2$}-norm minimization problem
\begin{equation}\label{eqn:ell2mini:single}
    \boldsymbol{\hat{\vartheta}} = \arg\min_{\boldsymbol{\vartheta}} \left\lVert\boldsymbol{r} - \boldsymbol{\Phi}\left(a_{1}\boldsymbol{\sin}_{\omega}+a_{2}\boldsymbol{\cos}_{\omega}\right)\right\rVert_2^2,
\end{equation}
where \smash{$\boldsymbol{\vartheta}=\left\lbrace\omega,a_{1},a_{2}\right\rbrace$} contains the reformulated sinusoid parameters. \eqref{eqn:ell2mini:single} can be rewritten in matrix form
\begin{equation}\label{eqn:ell2mini:matrix}
    \boldsymbol{\hat{\vartheta}} = \arg\min_{\boldsymbol{\vartheta}} \left\lVert\boldsymbol{r} - \boldsymbol{A}_{\omega}\boldsymbol{a}\right\rVert_2^2,
\end{equation}
where \smash{$\boldsymbol{A}_{\omega}=\left\lbrack\boldsymbol{\Phi}\boldsymbol{\sin}_{\omega},\boldsymbol{\Phi}\boldsymbol{\cos}_{\omega}\right\rbrack$} is composed of the compressed measurements of sinusoid samples at frequency \smash{$\omega$}, and \smash{$\boldsymbol{a}=\left\lbrack a_{1},a_{2}\right\rbrack^T$} contains the amplitudes of sine and cosine sinusoids, respectively. Given frequency \smash{$\omega$} \emph{a priori} known, \eqref{eqn:ell2mini:matrix} is converted to the estimation of amplitudes \smash{$\boldsymbol{a}$} by solving an \smash{$\ell_2$}-norm minimization problem
\begin{equation}\label{eqn:ell2mini:amp}
    \boldsymbol{\hat{a}} = \arg\min_{\boldsymbol{a}\in\mathbb{R}^2}\left\lVert\boldsymbol{r} - \boldsymbol{A}_{\omega}\boldsymbol{a}\right\rVert_2^2,
\end{equation}
to which the solution is given in a simple form \smash{$\boldsymbol{\hat{a}} = \boldsymbol{A}_{\omega}^{\dagger}\boldsymbol{r}$}, where \smash{$\boldsymbol{A}_{\omega}^{\dagger}=(\boldsymbol{A}_{\omega}^H\boldsymbol{A}_{\omega})^{-1}\boldsymbol{A}_{\omega}^H$} denotes the pseudoinverse of matrix \smash{$\boldsymbol{A}_{\omega}$}. Applying \eqref{eqn:ell2mini:amp} on \eqref{eqn:ell2mini:matrix} yields the following two-step optimization problem, which estimates frequency \smash{$\omega$} and amplitudes \smash{$\boldsymbol{a}$} consecutively.
\begin{equation}
    \hat{\omega}=\arg\min_{\omega\in\left\lbrack0,\pi\right\rbrack}\left\lVert\boldsymbol{r} - \boldsymbol{A}_{\omega}\boldsymbol{A}^{\dagger}_{\omega}\boldsymbol{r}\right\rVert_2^2,
    \boldsymbol{\hat{a}}=\boldsymbol{A}_{\hat{\omega}}^{\dagger}\boldsymbol{r}.
\end{equation}
\indent In Algorithm \ref{algorithm:parameter}, the estimation of frequency \smash{$\omega$} and amplitudes \smash{$\boldsymbol{a}$} are carried out in an iterative form to gradually converge to the numerical solution to \eqref{eqn:ell2mini:single}. In each iteration, the minimum \smash{$\ell_2$}-norm error is calculated for each frequency point sampled from predetermined frequency range with equal interval, and then, the frequency range is shrunk to the neighborhood of the frequency point with least \smash{$\ell_2$}-norm error, through which the estimation precision is improved.

\begin{algorithm}[tb!]
    \caption{Model Selection for Spectral Compressive Sensing}
    Input: CS matrix $\boldsymbol{\Phi} \in \mathbb{R}^{M \times N}$, CS measurement vector $\boldsymbol{m} \in \mathbb{R}^M$, sparsity $K$\\
    Output: Reconstructed frequency-sparse signal $\boldsymbol{\hat{s}}$\\
    Initialize: $\boldsymbol{\hat{s}}^{\left(j\right)}=0,\ j=1, \cdots,K$\\
    \While{halting criterion false}{
        \For{$i = 1\ \text{to}\ K$}{
            $\left\lbrace\text{form residual measurement}\right\rbrace$\\
            $\boldsymbol{r} \leftarrow \boldsymbol{m}-\sum_{\stackrel{j=1}{j\neq i}}^K \boldsymbol{\Phi}\boldsymbol{\hat{s}}^{\left(j\right)}$ \\
            $\left\lbrace\text{estimate sinusoid parameters}\right\rbrace$\\
            $\left\lbrace\hat{\omega}_i,\hat{a}_i,\hat{\phi}_i\right\rbrace \leftarrow \mathcal{R}\left(\boldsymbol{\Phi},\boldsymbol{r}\right)$\\
            $\left\lbrace\text{form sinusoid estimate}\right\rbrace$\\
            $\boldsymbol{\hat{s}}^{\left(i\right)} \leftarrow \left\lbrace \hat{a}_i \sin\left(\hat{\omega}_i t+\hat{\phi}_i\right)\right\rbrace_{t=1}^N$\\
        }
    }
    return $\boldsymbol{\hat{s}} \leftarrow \sum_{j=1}^K\boldsymbol{\hat{s}}^{\left(j\right)}$
    \label{algorithm:model}
\end{algorithm}

\begin{algorithm}[tb!]
    \caption{Sinusoid Parametric Estimation $\mathcal{R}\left(\boldsymbol{\Phi},\boldsymbol{r}\right)$\label{IR}}
    Input: CS matrix $\boldsymbol{\Phi}\in \mathbb{R}^{M \times N}$, residual CS measurement vector $\boldsymbol{r} \in \mathbb{R}^M$\\
    Output: Sinusoid parameter estimates $\hat{\omega}, \hat{a}, \hat{\phi}$\\
    Initialize: $\alpha=0,\ \beta=\pi,\ S=\infty$\\
    \While{halting criterion false}{
        $\left\lbrace\text{frequency estimate}\right\rbrace$\\
        $\left\lbrace\omega_i\right\rbrace_{i=0}^N \leftarrow \left\lbrace\alpha + i\left(\beta-\alpha\right)/N\right\rbrace_{i=0}^N$\\
        \For{$i = 0\ \text{to}\ N$}{
            $\left\lbrace\text{calculate compressed measurement}\right\rbrace$\\
            $\boldsymbol{A}_{\omega_i} \leftarrow \left\lbrack\boldsymbol{\Phi} \boldsymbol{\sin}_{\omega_i},\boldsymbol{\Phi} \boldsymbol{\cos}_{\omega_i}\right\rbrack$\\
            $\left\lbrace\text{calculate square error}\right\rbrace$\\
            $S_{\omega_i} \leftarrow \left\lVert \boldsymbol{r}- \boldsymbol{A}_{\omega_i}\boldsymbol{A}_{\omega_i}^{\dagger}\boldsymbol{r} \right\rVert_2^2$\\
            \If{$S_{\omega_i}<S$}{
                $S \leftarrow S_{\omega_i},j \leftarrow i$\\
                $\left\lbrace\text{update parameter estimate}\right\rbrace$\\
                $\boldsymbol{\hat{a}} = \left\lbrack\hat{a}_1,\hat{a}_2\right\rbrack^T \leftarrow \boldsymbol{A}_{\omega_i}^{\dagger}\boldsymbol{r},\ \hat{\omega} \leftarrow \omega_i $
            }
        }
        $\left\lbrace\text{frequency range refinement}\right\rbrace$\\
        $\alpha \leftarrow \max\left\lbrace\omega_{j-1},\alpha\right\rbrace,\beta \leftarrow \min\left\lbrace\omega_{j+1},\beta\right\rbrace$
    }
    return $\hat{\omega},\hat{a} \leftarrow \left(\hat{a}_1^2+\hat{a}_2^2\right)^{\frac{1}{2}},\hat{\phi} \leftarrow \arctan\left(\hat{a}_2/\hat{a}_1\right)$
    \label{algorithm:parameter}
\end{algorithm}

\section{Numerical Experiments}
\label{sec:num}
We compared the reconstruction performance of our algorithm, denoted as MDS, to state-of-the-art methods including \smash{$\ell_1$}-analysis, \smash{$\ell_1$}-synthesis, SIHT, BISP, BOMP, SDP and CBP, from both noisy and noise-free measurements\footnote{The authors would like to thank Marco F. Duarte, Karsten Fyhn, Boaz Nadler and Gongguo Tang (listed in alphabetical order) for providing the implementation of their algorithms.}. We chose the normalized $\ell_2$-norm error as major performance measure, which is defined as \smash{$\left\lVert x-\hat{x}\right\rVert_2/\left\lVert x\right\rVert_2$}, between the original signal \smash{$x$} and the recovered signal \smash{$\hat{x}$}. To evaluate and compare to other algorithms the frequency estimation performance of MDS, we generated frequency-sparse signals of length \smash{$N=128$} composed of \smash{$K=3$} real-valued sinusoids with frequencies selected uniformly at random, unit amplitudes, and zero phase shifts. The frequencies were well-separated so that no two tones were closer than \smash{$\pi/N$} to keep compatible with algorithms based on coherence inhibition signal model, including SIHT, BOMP, and BISP. Such separation is reasonable due to recent theoretical advances in relation between signal recoverability and minimum separation between spectral spikes\cite{tang2012compressed,candes2013towards}. Additionally, we generated frequency-sparse signals with frequencies as in the former setting but amplitudes and phase shifts selected uniformly at random to evaluate the parametric estimation performance of MDS. The two parameter settings are denoted in our experiment as MDS-FREQ and MDS-SINU, respectively. We performed Monte Carlo experiments and averaged over 600 trials. The sensing matrix \smash{$\boldsymbol{\Phi}$} was chosen as the Gaussian random matrix\footnote{For the SDP algorithm we used a random subsampling matrix, as the algorithm is only defined for such a sensing matrix.}, and the redundant DFT frame was with \smash{$c = 5$}.
\begin{figure}[tb!]
\includegraphics[width=0.485\textwidth]{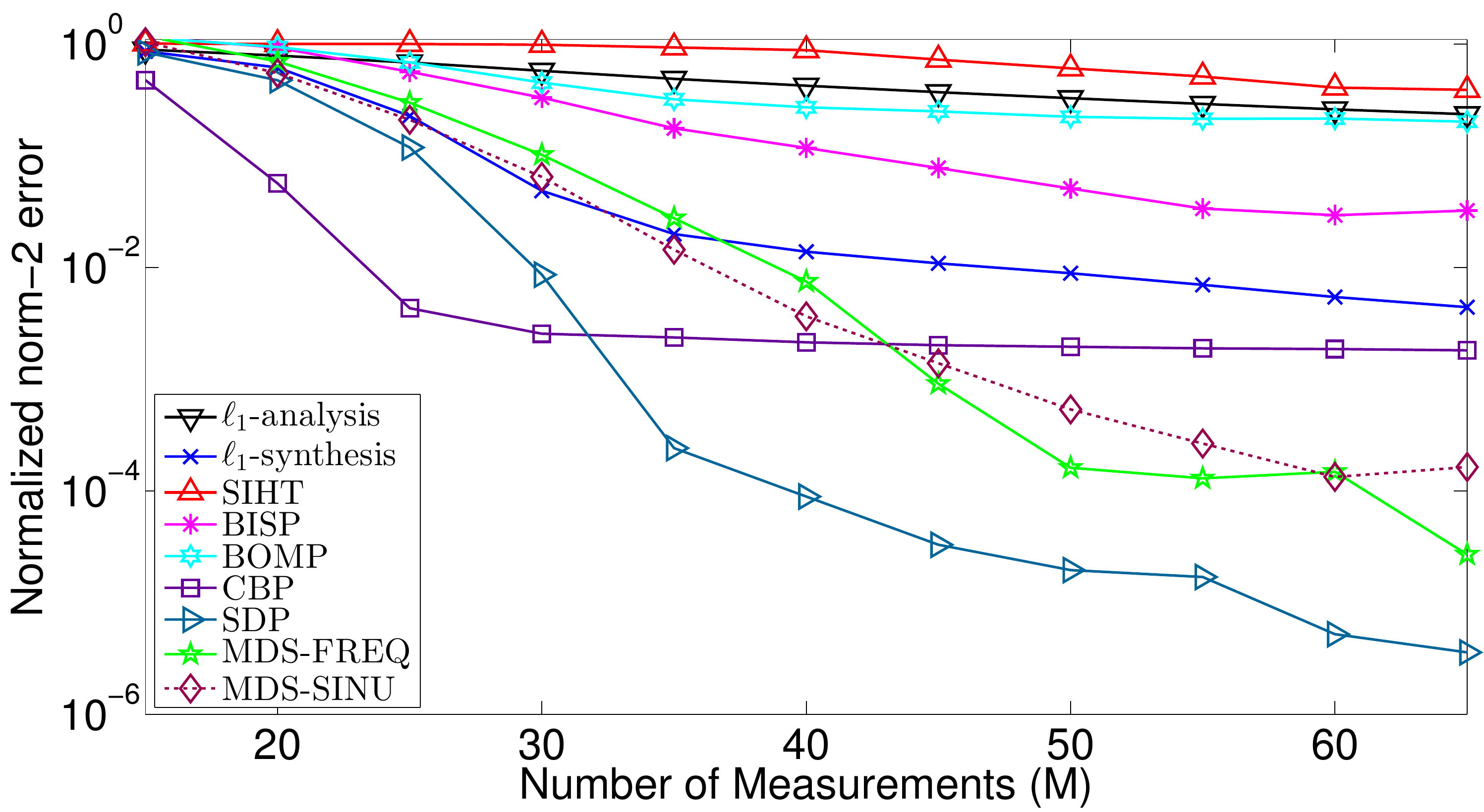}
\caption{Signal reconstruction performance in noiseless case.}
\label{figure:purereconstruct}
\end{figure}
\newline\indent In the first experiment, we evaluated the reconstruction performance from noiseless measurements with \smash{$M$} varying from 15 to 65. We set \smash{$\epsilon=10^{-10}$} for relevant algorithms. The results of numerical experiment are shown in Figure \ref{figure:purereconstruct}. In the noiseless case, SDP obtains the best result. When the number of measurements is sufficiently large, the frequency estimation performance of MDS outperforms CBP, whereas for M smaller than 40 it is worse than CBP and $\ell_1$-synthesis, while still better than other algorithms. The parametric estimation performance of MDS is similar to the frequency estimation. Among other algorithms, CBP outperforms \smash{$\ell_1$}-synthesis and remains static precision level for a wide range of M. Though the redundant DFT coefficients recovered by \smash{$\ell_1$}-synthesis is actually not sparse and exhibits severe frequency mismatch phenomenon, the signal is in practice reconstructed accurately. The performance of \smash{$\ell_1$}-analysis, BOMP, and SIHT is the worst among the algorithms tested.
\begin{figure}[tb!]
\centering
\includegraphics[width=0.485\textwidth]{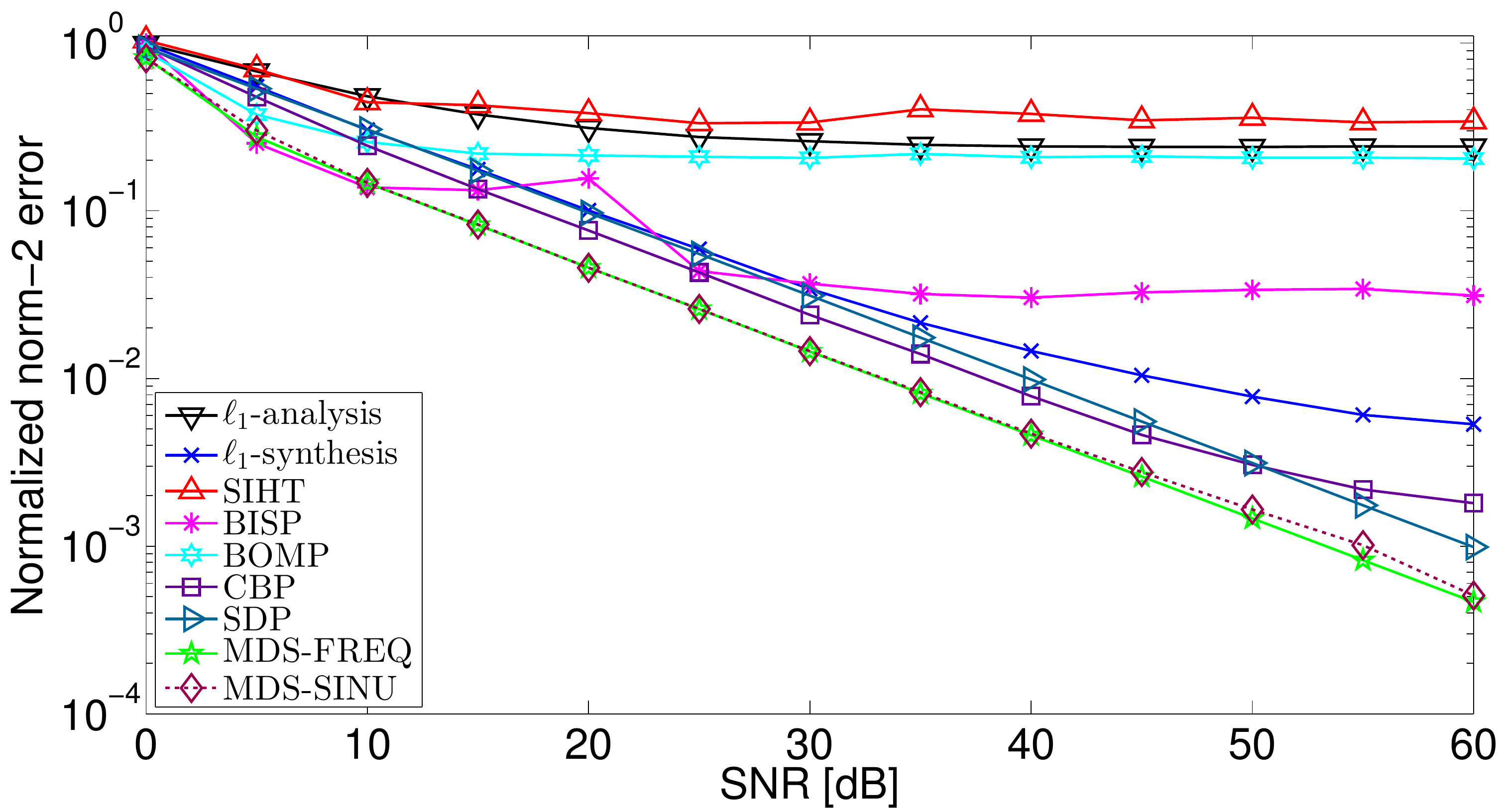}
\caption{Signal reconstruction performance in noisy case.}
\label{figure:noisereconstruct}
\end{figure}
\begin{table}[tb!]
\centering
\begin{tabular}{lrr}
    \hline
    Methods               & Noiseless & Noisy \\
    \hline
    $\ell_1$-analysis     & 232.5267 & 329.2279 \\
    $\ell_1$-synthesis    &   9.6257 &  12.8537 \\
    SIHT                  &   1.0950 &   1.1129 \\
    SDP                   &  39.2334 &  59.1094 \\
    BOMP                  &   0.0398 &   0.0396 \\
    CBP                   & 133.7628 & 115.8614 \\
    BISP                  &  17.7901 &  15.7219 \\
    MDS-FREQ              &   0.9900 &   0.9499 \\
    MDS-SINU              &   1.1507 &   1.0233 \\
    \hline
\end{tabular}
\caption{Average computation times in seconds.}
\label{table:averagetimes}
\end{table}
\newline\indent In the second experiment, we included additive Gaussian noise in the signal model. We fixed \smash{$M=64$} and explored a wide range of signal-to-noise ratio (SNR) value from 0 to 60 dB. The resulting performance curves are shown in Fig. \ref{figure:noisereconstruct}. In the noisy case, the frequency estimation performance of our algorithm outperforms other algorithms, and when random amplitudes and phase shifts are involved, only negligible deteriorations are induced. This is because model selection relies less on signal sparsity, and more on the matching to superposition of sinusoids. Among other algorithms, CBP and SDP obtain the best result, and \smash{$\ell_1$}-synthesis exhibits high fidelity when SNR level is sufficiently high. Despite its satisfactory performance in low SNR level, the performance of BISP is mediocre when noise level is low.
\newline\indent The computation time is of equal importance, and the average computation times are listed in Table \ref{table:averagetimes}\footnote{We set \smash{$M=64$} for the noiseless case and \smash{$\text{SNR}=30$} for the noisy case.}. The table shows that the excellent performance of frequency estimation and parameter estimation of MDS is enhanced by its high computational efficiency. Moreover, it is observed that the distinguished performance of CBP and SDP are restrained by their high computational expense. In addition, MDS only requires the sensing matrix to have the RIP. This flexibility on measurement scheme increases its performance advantage over SDP.

% Below is an example of how to insert images. Delete the ``\vspace'' line,
% uncomment the preceding line ``\centerline...'' and replace ``imageX.ps''
% with a suitable PostScript file name.
% -------------------------------------------------------------------------

% To start a new column (but not a new page) and help balance the last-page
% column length use \vfill\pagebreak.
% -------------------------------------------------------------------------
%\vfill
%\pagebreak

% References should be produced using the bibtex program from suitable
% BiBTeX files (here: strings, refs, manuals). The IEEEbib.bst bibliography
% style file from IEEE produces unsorted bibliography list.
% -------------------------------------------------------------------------
\newpage
\bibliographystyle{IEEEtran}
\bibliography{reference}

\end{document}